\documentclass[conference]{IEEEtran}
\IEEEoverridecommandlockouts

\usepackage{amsmath,amssymb,amsfonts}
\usepackage{algorithm}
\usepackage{algpseudocode}
\usepackage{graphicx}
\usepackage{textcomp}
\usepackage[dvipsnames]{xcolor}
\usepackage{booktabs}
\usepackage{array}
\usepackage{multirow}
\usepackage{url}
\usepackage{pifont}
\usepackage{marvosym} 
\usepackage{enumitem}
\usepackage[colorlinks=true,
            linkcolor=blue!55!black,
            citecolor=blue!55!black,
            urlcolor=blue!55!black,
            filecolor=blue!55!black,
            breaklinks=true,
            bookmarks=true,
            pdftitle={Database Context Compression for Text-to-SQL on Real-World Large Databases}]{hyperref}
\usepackage{cite}

\newcommand{\Sgcf}{\textnormal{\textsc{Sgcf}}}
\newcommand{\DbCC}{\textnormal{\textsc{DbCC}}}

\newtheorem{problem}{Problem}
\newtheorem{proposition}{Proposition}
\newtheorem{definition}{Definition}

\definecolor{upperformance}{RGB}{207,62,62}
\definecolor{downperformance}{RGB}{112,173,71}
\newcommand{\Up}[1]{\textcolor{upperformance}{#1\,$\uparrow$}}

\setlist[itemize]{leftmargin=*,itemsep=0pt,parsep=0pt,topsep=0pt,partopsep=0pt}
\setlist[enumerate]{leftmargin=*,itemsep=0pt,parsep=0pt,topsep=0pt,partopsep=0pt}

\begin{document}

\title{Database Context Compression for Text-to-SQL on Real-World Large Databases}

\author{%
\IEEEauthorblockN{%
Jingwen Liu\textsuperscript{1,2,3,*},
Weibin Liao\textsuperscript{1,2,3,*,\dag},
Xin Gao\textsuperscript{1,2,3},
Junfeng Zhao\textsuperscript{2,3},
Yasha Wang\textsuperscript{1,3,4,\,\Letter}%
\thanks{\textsuperscript{*}\,Equal contribution.\quad\textsuperscript{\dag}\,Project lead.\quad\textsuperscript{\Letter}\,Corresponding author.}}
\IEEEauthorblockA{%
\textsuperscript{1}National Engineering Research Center of Software Engineering, Peking University, Beijing, China\\
\textsuperscript{2}School of Computer Science, Peking University, Beijing, China\\
\textsuperscript{3}Key Laboratory of High Confidence Software Technologies, Ministry of Education, Beijing, China\\
\textsuperscript{4}Peking University Information Technology Institute, Tianjin Binhai, China\\
\texttt{liaoweibin@stu.pku.edu.cn}, \texttt{wangyasha@pku.edu.cn}}%
}

\maketitle

\begin{abstract}
Recent progress on Text-to-SQL has been driven by stronger language models and richer prompting strategies, yet performance on real enterprise benchmarks such as Spider~2.0 and BIRD remains far below that on classical academic datasets. We argue that the dominant bottleneck on such benchmarks is no longer reasoning ability, but the way the database is presented to the model. Real databases contain wide tables with repeated audit columns, large families of homogeneous partitioned tables, opaque machine-generated identifiers whose meaning lives only in column descriptions, and long data dictionaries in which only a small, query-dependent fraction is actually relevant. Existing query-aware approaches---schema linking, broadly construed to include retrieval-style schema subsetting---attempt to filter this raw context, but operate on top of a representation that is simultaneously structurally redundant, semantically verbose and documentation-heavy.

We re-frame the problem as one of \emph{database context compression}: a query-agnostic, database-side rewrite of the schema, semantic descriptors and external documents into a higher-density representation. We formalize this rewrite as the \Sgcf{} (Support--Gain Component Factorization) principle, which uniformly explains four very different operators---repeated column-group extraction, isomorphic table templating, shared semantic-tag componentization, and question-relevant evidence purification---as instances of a single coverage objective on different information layers. Building on \Sgcf{}, we present \DbCC{}, a two-phase database-side middleware that performs query-agnostic structural and semantic re-encoding offline, and lightweight query-aware evidence purification online. \DbCC{} is model-agnostic and pipeline-agnostic, and can be inserted before schema linking or generation in any existing Text-to-SQL system. On Spider~2.0-Snow and BIRD, \DbCC{} reduces input tokens by up to two orders of magnitude (on the most challenging \textsc{Large} bucket of Spider~2.0-Snow, $2.6\,\text{M}\!\to\!34.7$K tokens) while raising strict-recall on schema linking from $0\%$ to $56.5\%$ on that bucket under DeepSeek-V3.2 (and to $63.1\%$ under Claude-Opus-4.7), and yields a $1.8$--$1.9\%$ absolute end-to-end EX improvement when stacked on top of each of three recent Text-to-SQL systems. Our code is open-sourced at \url{https://github.com/MrBlankness/SchemaCompression}.
\end{abstract}

\begin{IEEEkeywords}
Text-to-SQL, large databases, context compression, schema linking, natural language interfaces to databases
\end{IEEEkeywords}

\begin{figure}[!t]
\centering
\includegraphics[width=\columnwidth]{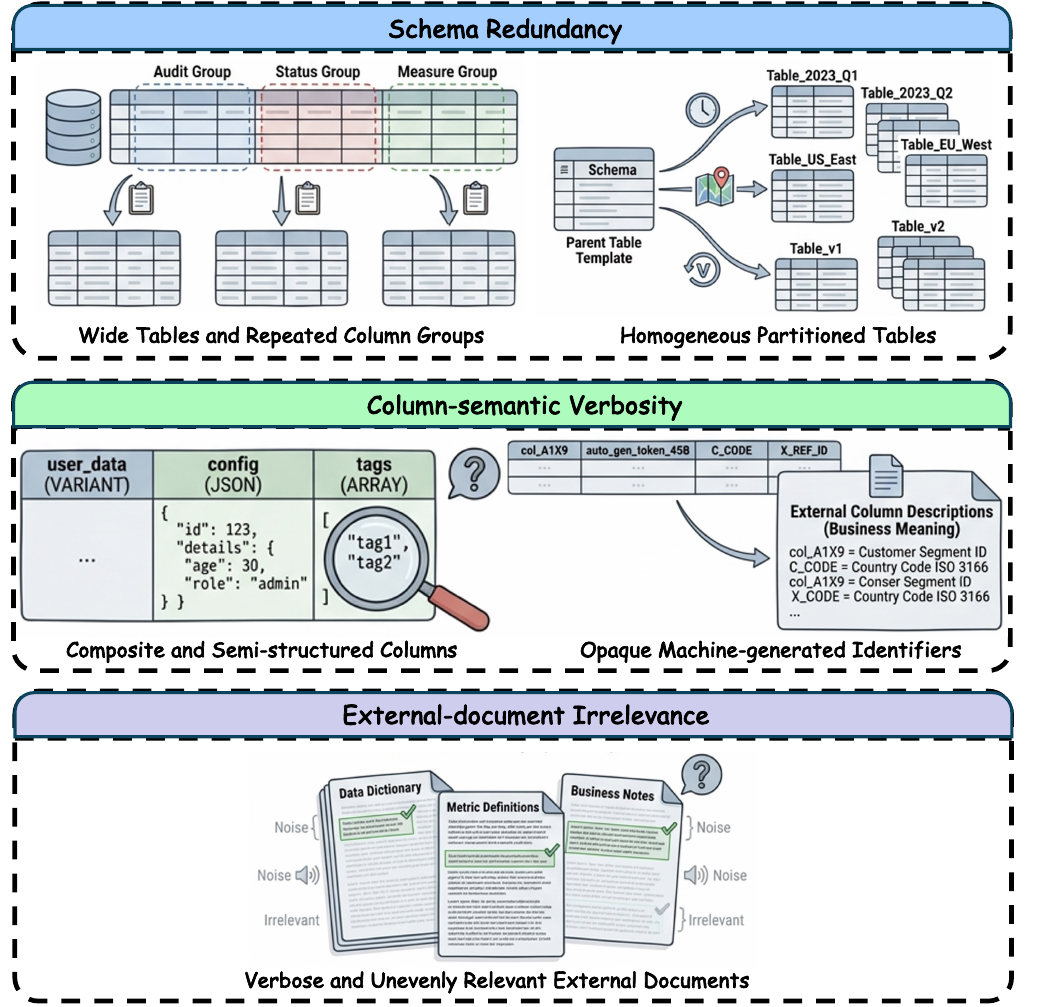}
\caption{The bottleneck of Text-to-SQL on real enterprise databases. A representative Spider~2.0 schema contains thousands of columns drawn from wide audit-heavy tables, large families of sharded partitions, opaque machine-generated identifiers and lengthy data dictionaries; existing query-aware schema-linking pipelines must filter this raw context, but the input itself is structurally redundant, semantically verbose and documentation-heavy. \DbCC{} takes the orthogonal path of \emph{rewriting the database} once on the database side---yielding a compressed view $D'(q)=(S',M',E'(q))$ that is then consumed by any downstream pipeline---instead of repeatedly filtering the raw $D$ per query.}
\label{fig:teaser}
\end{figure}

\section{Introduction}
\label{sec:intro}

Natural-language interfaces to relational databases promise to lower the bar for accessing the data that increasingly drives modern enterprises. The Text-to-SQL task, which translates a user question into an executable SQL query, is the workhorse of this vision~\cite{nalir,sqlizer,spider,athena,templar,duoquest,survey25a,survey25b}. Driven by large language models (LLMs), Text-to-SQL accuracy on classical academic benchmarks has reached impressive levels~\cite{ratsql,resdsql,dinsql,chess,chasesql,dailsql}.

The picture changes sharply once these systems are deployed on real enterprise data. Benchmarks such as BIRD~\cite{bird} and especially Spider~2.0~\cite{spider2} expose a substantial gap: average databases contain on the order of $10^3$ columns, queries involve long join chains, business semantics are often hidden in opaque identifiers, and the surrounding documentation may run to tens of pages per database. On such inputs, even the strongest LLMs degrade markedly. The natural reaction is to attribute this degradation to insufficient model capacity, longer reasoning chains, or richer prompts. Surveys of the field~\cite{survey25a,survey25b} have begun to suggest, however, that the bottleneck has shifted: the dominant source of error is no longer how the model reasons about the database, but how the database is presented to the model in the first place. Fig.~\ref{fig:teaser} summarizes this shift: existing query-aware schema-linking pipelines are forced to filter a raw context that is itself structurally redundant, semantically verbose and documentation-heavy, whereas \DbCC{} \emph{rewrites the database} once on the database side and lets every downstream pipeline consume the resulting compact view.

\subsection{Why ``larger'' databases are not just larger}
\label{sec:intro-largeness}

Treating large databases as a quantitative scaling problem misses what makes them qualitatively different. Inspecting the databases shipped with Spider~2.0~\cite{spider2} and BIRD~\cite{bird} reveals that the growth of physical schema size is dominated by \emph{engineering artefacts} rather than business semantics:

\begin{itemize}
\item \textbf{Wide tables and repeated column groups.} For performance and historical reasons, business tables are often denormalized, and identical groups of audit, status or measurement columns are pasted across many tables.
\item \textbf{Homogeneous partitioned tables.} Time, region or version sharding produces large families of physical tables whose schemas are essentially identical to a common parent.
\item \textbf{Composite and semi-structured columns.} \texttt{VARIANT}, \texttt{JSON}, \texttt{ARRAY} and \texttt{STRUCT} columns hide the discriminative information inside nested attributes, where neither column name nor type alone is sufficient for matching.
\item \textbf{Opaque machine-generated identifiers.} Many column names are abbreviations, codes or auto-generated tokens whose business meaning lives only in external column descriptions.
\item \textbf{Verbose and unevenly relevant external documents.} Data dictionaries, metric definitions and business notes provide essential disambiguation, but each document is long, only a small fraction is relevant to any given question, and the rest acts as in-context noise.
\end{itemize}

These five observations are not five independent problems. They cluster into three challenges, each defined by the layer of the database context at which low information density arises and by whether that density is intrinsic to the database or to the question.

\textbf{Challenge\,\#1 (Schema redundancy).} Observations~1--2 inflate the \emph{schema} layer with content that is repeated across tables: identical groups of audit, status or measurement columns reappear in many wide tables, and isomorphic schemas reappear across families of sharded tables. The redundancy is structural and query-agnostic---a column group repeated in 30 tables is repeated regardless of who is asking.

\textbf{Challenge\,\#2 (Column-semantic verbosity).} Observations~3--4 inflate the \emph{column-semantics} layer: composite and semi-structured columns force the system to inspect nested attributes that neither column name nor type alone disambiguates, while opaque machine-generated identifiers shift the burden of meaning onto verbose external column descriptions. The cost paid per column is high, and---because similar concepts are described differently across tables---the same semantic content is paid for many times.

\textbf{Challenge\,\#3 (External-document irrelevance).} Observation~5 inflates the \emph{external-knowledge} layer in a qualitatively different way: each individual document is internally non-redundant, but only a small, question-dependent fraction of any document is relevant to a given question; the remainder acts as in-context noise. Unlike Challenges\,\#1--\#2, the redundancy here cannot be removed once and for all---it depends on the question.

The combined effect is that real database context is simultaneously \emph{too large to fit} and \emph{too unfocused to use}. Sec.~\ref{sec:framework} returns to these three challenges and shows how each is addressed by a corresponding operator inside \DbCC{}.

\subsection{The shift we propose: from query-aware filtering to query-agnostic re-encoding}
\label{sec:intro-shift}

Existing approaches to large-database Text-to-SQL implicitly assume that the raw database representation is fundamentally adequate, and that the system's job is to filter, rank or select within it. Schema-linking methods, broadly construed---ranging from finer-grained alignment over the original schema (\textsc{Rat-SQL}~\cite{ratsql}, \textsc{ResdSQL}~\cite{resdsql}, \textsc{Bridge}~\cite{bridge}, \textsc{RSL-SQL}~\cite{rslsql}) to retrieval-style subsetting (\textsc{Crush4SQL}~\cite{crush4sql}, \textsc{LinkAlign}~\cite{linkalign}, \textsc{CodeS}~\cite{codes}, \textsc{DB-Explore}~\cite{dbexplore})---all decide what to keep on top of an unchanged database; agentic methods such as \textsc{Chess}~\cite{chess}, \textsc{ReFoRCE}~\cite{reforce}, \textsc{Mac-SQL}~\cite{macsql} and \textsc{AutoLink}~\cite{autolink} explore the database iteratively, but again over its raw form. None of them rewrites the database itself.

We argue that this is precisely what should be rewritten. The redundancy in real schemas is structural and largely query-agnostic, and the verbosity of column descriptions and external documents is intrinsic to how these artefacts were produced rather than to the question. Treating these regularities as engineering noise that should be \emph{compressed at the database side, once and for all} aligns with classical database thinking---one builds an index, a materialized view or a summary statistic so that \emph{every} subsequent query benefits. The re-framing has three direct consequences: (i)~it separates a query-agnostic offline stage from the lightweight, query-aware online stage of evidence purification; (ii)~it makes the offline stage a reusable database-side asset whose cost amortizes over many queries; (iii)~it is orthogonal to query-aware methods, providing them with a more compact, more focused input.

\subsection{Contributions}
\label{sec:intro-contrib}

This paper makes four contributions.

\begin{itemize}
\item \textbf{A new framing.} We identify \emph{database context compression} as a first-class problem layer for Text-to-SQL on real databases, distinct from schema linking and prompt compression (Sec.~\ref{sec:formal}).
\item \textbf{A unified principle.} We introduce the \Sgcf{} principle (Support--Gain Component Factorization), a single coverage-style objective that uniformly captures structural, semantic and external-knowledge compression as instances of the same rewrite operator on different information layers (Sec.~\ref{sec:framework}).
\item \textbf{Concrete algorithms.} We design and analyze four operators under \Sgcf{}: column-group factorization for wide tables (Sec.~\ref{sec:struct}), template hierarchies for homogeneous partitions (Sec.~\ref{sec:struct}), shared semantic-tag componentization for column descriptors (Sec.~\ref{sec:semantic}), and question-relevant evidence purification for external knowledge (Sec.~\ref{sec:external}).
\item \textbf{A pluggable middleware.} We package the operators into \DbCC{}, a two-phase database-side middleware that integrates with any Text-to-SQL system---traditional pipelines, prompt-based LLMs and agentic flows---without modifying their core logic (Sec.~\ref{sec:integration}).
\end{itemize}

The remainder of the paper is organized as follows. Section~\ref{sec:related} relates \DbCC{} to the most closely connected lines of work. Section~\ref{sec:formal} formalizes database context compression. Section~\ref{sec:framework} presents the \Sgcf{} principle and the \DbCC{} architecture. Sections~\ref{sec:struct}--\ref{sec:external} describe the structural, semantic and external-knowledge operators. Section~\ref{sec:integration} discusses integration with downstream systems. Section~\ref{sec:exp} reports the experimental evaluation. Section~\ref{sec:conclusion} concludes.

\section{Related Work}
\label{sec:related}

We position \DbCC{} against three lines of work and explicitly mark the boundary at each line.

\subsection{Text-to-SQL with LLMs}

The transition from supervised parsers~\cite{ratsql,resdsql,lgesql,grappa,strug,bridge,picard,irnet} to LLM-based pipelines~\cite{c3sql,dinsql,actsql,dailsql,petsql,chess,chasesql,purple,aidsql} has changed the focus of the field from designing a stronger parser to organizing a more useful task context. Methods such as \textsc{Din-SQL}~\cite{dinsql}, \textsc{Chess}~\cite{chess} and \textsc{Chase-SQL}~\cite{chasesql} decompose the prompt or aggregate multiple reasoning paths; \textsc{Mac-SQL}~\cite{macsql}, \textsc{ReFoRCE}~\cite{reforce} and \textsc{AutoLink}~\cite{autolink} generalize this view to multi-agent or self-refining flows; generate-then-rank pipelines such as \textsc{Gar}~\cite{gar} and \textsc{Metasql}~\cite{metasql} rerank an SQL-candidate pool---rather than schema candidates---against the input question. Surveys of LLM-era Text-to-SQL~\cite{survey25a,survey25b}, in-depth benchmarking studies~\cite{indepthbench}, and benchmark papers such as Spider~2.0~\cite{spider2} and BIRD~\cite{bird} document a clear pattern: improvements at the model side saturate, while context-side issues---noise, redundancy and length---become the new dominant factor. \DbCC{} sits squarely on the database side of this divide and is complementary to all of the above.

\subsection{Schema linking}

Schema linking---understood broadly as deciding which elements of the database are relevant to a question---spans two complementary granularities. Token-level alignment was first developed inside end-to-end parsers (\textsc{Rat-SQL}~\cite{ratsql}, \textsc{ResdSQL}~\cite{resdsql}, \textsc{Bridge}~\cite{bridge}) and later isolated as a standalone module in dedicated linkers (\textsc{RSL-SQL}~\cite{rslsql}, \textsc{LinkAlign}~\cite{linkalign}, \textsc{ISESL-SQL}~\cite{isesl}); retrieval-style approaches~\cite{crush4sql,codes,dbexplore,schemamatchplm} narrow a far larger candidate space when the full schema cannot be fed in directly. Disambiguation frameworks such as \textsc{CLEAR}~\cite{clearnl2sql} target a different axis---resolving question-side ambiguity rather than schema-side redundancy---and are likewise complementary to \DbCC{}. Both families are \emph{query-aware}: they decide what is relevant given the user question, on top of an unchanged database representation. \DbCC{} differs in two crucial ways. First, it is \emph{query-agnostic} for its structural and semantic operators: it rewrites the database once, and the rewrite serves every subsequent query. Second, it is \emph{lossless with respect to schema content}: a templated table or a factorized column group can always be expanded back to its original definition, so downstream linking does not lose any candidates. In our integration analysis (Sec.~\ref{sec:integration}) and ablations (Sec.~\ref{sec:exp}) we therefore evaluate \DbCC{} both standalone and stacked on top of representative schema-linking baselines.

\subsection{Prompt compression and schema description}

Generic prompt-compression methods such as \textsc{LLMLingua}~\cite{llmlingua} and the \emph{Lost-in-the-Middle} analysis~\cite{lostmiddle} target token-level redundancy in arbitrary prompts. \textsc{Schemonic}~\cite{schemonic} is closest in spirit to our work: it generates succinct natural-language descriptions of database schemas to lower the cost of LLM prompting. A separate line of database-side adaptations targets specific deployment regimes---domain-specific fine-tuning frameworks such as \textsc{FinSQL}~\cite{finsql}, hybrid small/large LM compositions~\cite{smallllmnl2sql}, and very-long-context LLM strategies~\cite{longcontextnl2sql}---all of which still consume the raw schema verbatim. \DbCC{} differs in that we exploit \emph{structural} regularities of the database (column-group repetition, table isomorphism, semantic-tag sharing) to perform a \emph{lossless} rewrite, rather than a textual summarization or an enlarged context window. The two are complementary: \textsc{Schemonic}-style summaries can be applied to the residual textual layer that survives our structural compression.

\section{Problem Formalization}
\label{sec:formal}

This section formalizes \emph{database context compression} as a database-side rewrite problem.

\subsection{Database context}

Let a database $\mathcal{D}$ be associated with three kinds of context information that a Text-to-SQL system may consume:

\begin{itemize}
\item the \textbf{schema} $S$, comprising tables and their typed columns and key relationships;
\item the \textbf{column-level semantics} $M$, comprising natural-language column descriptions, types and sample values;
\item the \textbf{external knowledge} $E$, comprising data dictionaries, metric definitions and business documents.
\end{itemize}

The raw database context is denoted $D=(S,M,E)$. Given a question $q$, an end-to-end Text-to-SQL system $\Pi$ produces a SQL query
\[
\hat{y} = \Pi\bigl(q, D\bigr).
\]

\subsection{The compression problem}

A database compressor is a map
\[
\mathcal{C}:\; D \;\longmapsto\; D'(q) = (S', M', E'(q))
\]
that rewrites database context into a more compact representation $D'$, possibly conditioned on $q$ for the external-knowledge component. Let $\mathrm{Cost}(\cdot)$ denote a measure of input cost---throughout this paper we use \emph{input-token count}, which is also the dominant per-call billing unit for commercial LLM APIs---and $\mathrm{Util}(\Pi, D, \cdot)$ denote a measure of downstream utility (e.g., schema-linking recall or execution accuracy). The compressor is required to satisfy:

\begin{problem}[Database context compression]
\label{problem:dbcc}
Find $\mathcal{C}$ such that for a target downstream system~$\Pi$ and a workload of questions $\{q_i\}$, the compressed context simultaneously satisfies, \emph{in expectation over $\{q_i\}$},
\begin{equation}
\mathrm{Cost}\!\bigl(D'\bigr) \;\ll\; \mathrm{Cost}\!\bigl(D\bigr)
\label{eq:problem-cost}
\end{equation}
and
\begin{equation}
\mathrm{Util}\!\bigl(\Pi, D', q_i\bigr) \;\ge\; \mathrm{Util}\!\bigl(\Pi, D, q_i\bigr).
\label{eq:problem-util}
\end{equation}
\end{problem}

Three properties of Problem~\ref{problem:dbcc} are worth emphasizing.

\textbf{(i) Asymmetric in $q$.} The structural part $S'$ and the semantic part $M'$ are produced by query-agnostic rewrites: they depend only on $D$ and can be precomputed and cached at the database side. The external-knowledge part $E'$ is the only query-aware component and is computed online. This asymmetry is the basis for the two-phase architecture in Sec.~\ref{sec:framework}.

\textbf{(ii) Lossless with respect to schema content.} \DbCC{} discards no base table or column. Compression is realized through \emph{structured references}---shared components and template inheritance---that can be deterministically expanded back, so that downstream linking has access to the full candidate space whenever needed.

\textbf{(iii) Inequality in utility, not equality.} Equation~\eqref{eq:problem-util} permits strict improvement, $\mathrm{Util}(D')\ge\mathrm{Util}(D)$, because compressing redundant or noisy context often \emph{improves} downstream accuracy by removing distractors---a phenomenon consistent with the \emph{Lost-in-the-Middle} effect~\cite{lostmiddle}.

\section{The \DbCC{} Framework}
\label{sec:framework}

\begin{figure*}[!t]
\centering
\includegraphics[width=0.8\textwidth]{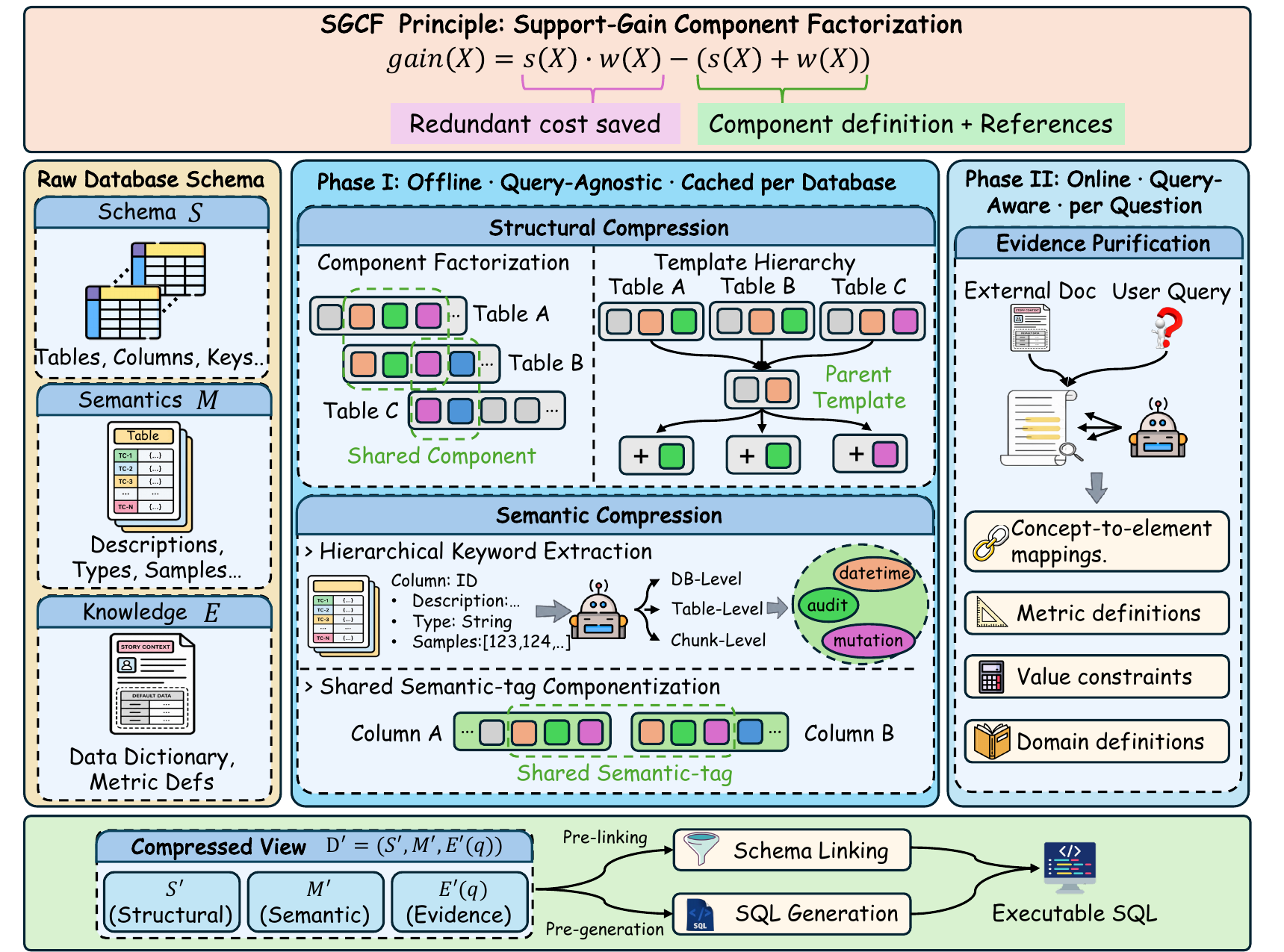}
\caption{Architecture of \DbCC{}. \textbf{Phase~I (offline, query-agnostic, per database)} rewrites the raw database context $D=(S,M,E)$ into a compact view through two operators: a \emph{structural} operator that turns wide tables and families of sharded tables into column-group factorizations and template hierarchies $S\!\to\!S'$, and a \emph{semantic} operator that turns verbose column descriptions into hierarchical keyword tags and shared-tag components $M\!\to\!M'$. Phase~I outputs a cached view $(S',M')$ that is reused by \emph{every} subsequent query. \textbf{Phase~II (online, query-aware, per question)} runs only when a question $q$ arrives: the \emph{evidence-purification} operator turns long external documents $E$ into a small set of typed evidence atoms $E'(q)$ aligned to the schema. The final output $D'(q)=(S',M',E'(q))$ is consumed without modification by downstream Text-to-SQL systems---schema-linking pipelines, prompt-based LLMs or agentic flows. The two-phase split makes the dominant LLM-driven cost (semantic keyword extraction in Phase~I) amortize across queries, while keeping per-query work to a thin question-conditioned step.}
\label{fig:arch}
\end{figure*}

This section shows how Problem~\ref{problem:dbcc} is solved uniformly across very different information layers.

\subsection{The \Sgcf{} principle}
\label{sec:sgcf}

The redundancies described in Sec.~\ref{sec:intro-largeness}---repeated column groups across wide tables, isomorphic families of partitioned tables, repeated semantic tags across column descriptions, and repeated business concepts scattered across external documents---all share the same structure. They are \emph{repeated subsets of information units carried by multiple objects}.

\begin{definition}[Information layer]
\label{def:layer}
An information layer is a triple $\langle\mathcal{O}, U, \mathrm{cost}\rangle$ where $\mathcal{O}$ is a set of \emph{carrier objects} (e.g., tables, columns, document chunks), $U:\mathcal{O}\to 2^{\mathcal{V}}$ assigns to each carrier a set of \emph{information units} drawn from a vocabulary $\mathcal{V}$, and $\mathrm{cost}(\cdot)$ is a per-unit cost function (e.g., tokens).
\end{definition}

A \emph{component} on an information layer is a subset $X\subseteq\mathcal{V}$ of information units shared by multiple carriers. Replacing every shared occurrence of $X$ with a reference to a single component definition removes redundancy.

\begin{definition}[Support, width and gain]
\label{def:gain}
For a candidate component $X\subseteq\mathcal{V}$ on layer $\langle\mathcal{O}, U, \mathrm{cost}\rangle$, its \emph{support} $s(X)$ is the number of carriers that contain $X$, its \emph{width} $w(X)$ is the cardinality of $X$, and its \emph{gain} is
\[
s(X) = \bigl|\{o\in\mathcal{O} : X\subseteq U(o)\}\bigr|,
\qquad w(X) = |X|,
\]
\begin{equation}
\mathrm{gain}(X) \;=\; \underbrace{s(X)\cdot w(X)}_{\text{redundant cost saved}} - \underbrace{\bigl(s(X)+w(X)\bigr)}_{\text{component definition + references}},
\label{eq:gain}
\end{equation}
under the convention that each information unit and each reference token costs one unit (the absolute scale of $\mathrm{gain}$ thus differs from real tokenizer costs by a small constant factor, but the ranking of components is preserved).
\end{definition}

\begin{proposition}[Unified compression objective]
\label{prop:unified}
Let $\mathcal{X}$ be a family of admissible components subject to layer-specific feasibility constraints (e.g., minimum support, maximum width, no overlapping with already-selected components). The unified compression problem is to maximize the total gain
\begin{equation}
\max_{\mathcal{X}}\;\sum_{X\in\mathcal{X}}\mathrm{gain}(X),
\label{eq:objective}
\end{equation}
which is a coverage-style objective with a per-component reward $s(X)\!\cdot\!w(X)$ and a per-component definition penalty $s(X)\!+\!w(X)$. \DbCC{} solves it on each layer with a greedy heuristic that iteratively selects the highest-gain admissible component (Algorithms~\ref{alg:fact}--\ref{alg:freq}); we treat this as a layer-specific design choice rather than a worst-case approximation guarantee, since the penalty term breaks the standard non-negative monotone-submodular setup that would otherwise yield a $(1\!-\!1/e)$ bound.
\end{proposition}

What is novel is the proposition's \emph{unifying role} across very different layers. Every operator inside \DbCC{} (Sec.~\ref{sec:struct}--\ref{sec:external}) instantiates Equation~(\ref{eq:objective}) with a different carrier/information-unit pair: tables and column signatures (horizontal structural), table clusters and column sets (vertical structural), columns and keyword sets (semantic), document chunks and evidence-span sets (external knowledge). The three layers are in one-to-one correspondence with the challenges of Sec.~\ref{sec:intro-largeness}: Challenge\,\#1 (schema redundancy) is targeted by the structural layer, with column-group factorization (Sec.~\ref{sec:struct-fact}) handling the \emph{horizontal} variant from wide tables and template hierarchy (Sec.~\ref{sec:struct-inh}) the \emph{vertical} variant from homogeneous partitions; Challenge\,\#2 (column-semantic verbosity) is targeted by the semantic layer, where hierarchical keyword extraction normalizes per-column descriptions and shared semantic-tag componentization removes the resulting cross-column repetition (Sec.~\ref{sec:semantic}); Challenge\,\#3 (external-document irrelevance) is intrinsically question-specific and is targeted by question-driven evidence purification (Sec.~\ref{sec:external}). The first two layers are query-agnostic and can be precomputed once per database; the third is query-aware and is computed per question---an asymmetry that motivates the two-phase architecture introduced next.

\subsection{Architecture}

\DbCC{} realizes \Sgcf{} in a two-phase architecture (Fig.~\ref{fig:arch}). The split has four direct consequences: (C1)~\emph{Amortization}---the dominant LLM-driven step (semantic keyword extraction) lives in Phase~I, so its cost is paid once per database; (C2)~\emph{Reusability}---Phase~I produces a self-contained compressed view consumable by any downstream pipeline without modification; (C3)~\emph{Statefulness vs.\ statelessness}---Phase~I is a stateful database asset (akin to an index or materialized view), Phase~II is stateless and lightweight; (C4)~\emph{Compositionality}---because the three operators act on disjoint information layers, their gains are largely orthogonal, supporting clean ablation. A natural concern is whether structural, semantic and external-knowledge compression could cannibalize each other. Equation~(\ref{eq:gain}) makes the answer transparent: each operator selects components from a disjoint vocabulary $\mathcal{V}$ (column signatures vs.\ keyword tokens vs.\ evidence spans), so components selected by one operator cannot be selected by another. \DbCC{} quantifies orthogonality empirically with the index $\mathrm{Orth}(A,B) = \bigl(\Delta(A\cup B) - \max(\Delta A,\Delta B)\bigr)/\min(\Delta A,\Delta B)$, where $\Delta$ is the gain in downstream utility relative to the uncompressed baseline; positive values indicate that the two operators contribute additively (Sec.~\ref{sec:exp}).

\section{Structural Compression}
\label{sec:struct}

The structural layer addresses redundancy in the schema itself. \DbCC{} instantiates two operators that together cover the two dominant redundancy patterns of real schemas.

\subsection{Component factorization for wide tables}
\label{sec:struct-fact}

\paragraph*{Pattern} A \emph{column group} is a set of column signatures $X\subseteq\mathcal{V}_S$ that appears in many tables. Examples include audit columns (\texttt{created\_at}, \texttt{updated\_at}, \texttt{updated\_by}), monetary status columns (\texttt{amount}, \texttt{currency}, \texttt{status}), and standardized location attributes. Repeating $X$ in every table that contains it costs $|X|\cdot s(X)$ tokens; expressing $X$ once as a shared component and inserting a reference in every supporting table costs $|X|+s(X)$.

\paragraph*{Operator} For each candidate column group $X$, \DbCC{} computes its support $s(X)$, width $w(X)=|X|$ and gain~\eqref{eq:gain}, and accepts $X$ as a component iff $s(X)\ge\tau_s$, $w(X)\ge\tau_w$ and $\mathrm{gain}(X)\ge\tau_g$. To keep each table's rendering readable when many components apply, we additionally enforce a \emph{per-table component quota} $K$ (the maximum number of references attached to any single table; we use $K=8$ throughout).

\paragraph*{Algorithm} Naive enumeration of column subsets within a wide table is exponential. \DbCC{} avoids this with an inverted-index trick that exploits the structure of column groups.

\begin{algorithm}[t]
\caption{Column-group factorization}
\label{alg:fact}
\begin{algorithmic}[1]
\State $\phi(c) \!\leftarrow\! \{T \!\in\! \mathcal{T} : c \!\in\! T\}$ \quad for every column signature $c$.
\State $\mathcal{G} \!\leftarrow\! \{\, X_S \!=\! \phi^{-1}(S) : S\!\subseteq\!\mathcal{T}\,\}$ \quad (one candidate per equivalence class of $\phi$; see below).
\State $\mathcal{G}^\star \!\leftarrow\! \{\, X \!\in\! \mathcal{G} : s(X)\!\ge\!\tau_s,\; w(X)\!\ge\!\tau_w,\; \mathrm{gain}(X)\!\ge\!\tau_g \,\}$.
\State Sort $\mathcal{G}^\star$ by $(\mathrm{gain}(X), s(X), w(X))$ in decreasing order.
\State $\mathcal{C}\!\leftarrow\!\emptyset$.
\For{$X\!\in\!\mathcal{G}^\star$ in sorted order}
  \If{$X$ disjoint from every $Y\!\in\!\mathcal{C}$ \textbf{and} per-table quota $K$ respected}
    \State $\mathcal{C}\!\leftarrow\!\mathcal{C}\!\cup\!\{X\}$.
  \EndIf
\EndFor
\State \textbf{Output:} $S' \!=\! \{\,T \mapsto (\mathcal{R}(T),\, U(T)\!\setminus\!\bigcup_{X\in\mathcal{R}(T)} X)\,\}_{T\in\mathcal{T}}$, with $\mathcal{R}(T)\!=\!\{X\!\in\!\mathcal{C}: X\!\subseteq\! U(T)\}$.
\end{algorithmic}
\end{algorithm}

\paragraph*{Why the inverted index} A wide table can have hundreds of columns, so subset enumeration is intractable. The key observation is that two columns sharing the same support set $\phi(c)\!=\!\phi(c')$ must co-occur in any column group, so the only candidates worth scoring are the \emph{equivalence classes of $\phi$} taken whole. We thus restrict $\mathcal{G}$ to one candidate per distinct support set rather than enumerating all subsets of each class---a deliberate heuristic pruning that may miss strict subsets but in practice does not, because $\mathrm{gain}$~\eqref{eq:gain} is monotonically maximised by the widest co-occurring group at a given support. Components are rewritten as references, not destroyed, so the operator is reversible: any downstream component can request the expanded form.

\subsection{Template hierarchy for homogeneous partitions}
\label{sec:struct-inh}

\paragraph*{Pattern} A \emph{template} is a set of physical tables $\{T_1,\ldots,T_k\}$ whose column-set is identical. Time-, region- or version-sharded tables are templates by construction. A \emph{derived template} is a template whose column-set is a superset of another template's column-set, differing only by a small delta.

\paragraph*{Operator} \DbCC{} compresses a template family by recording the parent template once, plus, for each derived template, only the delta. Inheritance gives a compact tree-of-tables representation that resembles object-oriented inheritance.

\begin{algorithm}[t]
\caption{Template hierarchy construction}
\label{alg:inh}
\begin{algorithmic}[1]
\State $\mathcal{B}\!\leftarrow\!\{\,B_i\,\}$, where $B_i$ groups physical tables sharing the same column signature $\mathrm{Sig}(B_i)$.
\State Sort $\mathcal{B}$ by $|\mathrm{Sig}(\cdot)|$ in increasing order. \textbf{Process $B_i$ in this order}, so that any candidate parent $B_j$ with $\mathrm{Sig}(B_j)\!\subseteq\!\mathrm{Sig}(B_i)$ has already been admitted (or rejected) by the time $B_i$ is visited; when no such $B_j$ exists, $B_i$ is left as a root.
\State For each $B_i$, define
       \[
        \mathrm{Par}(B_i) = \arg\max_{B_j\,:\,\mathrm{Sig}(B_j)\subseteq\mathrm{Sig}(B_i),\;|B_j|\ge\tau_p}\;\bigl|\mathrm{Sig}(B_j)\bigr|,
       \]
       and $\Delta_i \!=\! \mathrm{Sig}(B_i)\!\setminus\!\mathrm{Sig}\bigl(\mathrm{Par}(B_i)\bigr)$.
\State Accept the inheritance link $B_i\!\to\!\mathrm{Par}(B_i)$ iff $\mathrm{gain}(B_i)\!\ge\!\tau_g$ and $|\Delta_i|/|\mathrm{Sig}(B_i)|\!\le\!\tau_r$; otherwise mark $B_i$ as a root.
\State \textbf{Output:} forest $\mathcal{F}=\{(B_i,\mathrm{Par}(B_i),\Delta_i)\}$ of accepted links plus the roots.
\end{algorithmic}
\end{algorithm}

\paragraph*{Design rationale and complementarity with factorization} Two thresholds control the rendering: the minimum parent size $\tau_p$ ensures that the parent carries enough structural information to be informative, and the relative-delta threshold $\tau_r$ rejects derivations whose increment is too large to read as a small extension---in pilot studies, removing $\tau_r$ produced ``inheritance'' relations that were technically valid but unreadable, hurting downstream linking despite reducing tokens. Factorization handles \emph{horizontal} redundancy (columns shared across tables) while template inheritance handles \emph{vertical} redundancy (entire tables sharing the same shape); the two operate on different carriers and select from disjoint candidate spaces, and their interaction is evaluated in Sec.~\ref{sec:exp}.

\section{Semantic Compression}
\label{sec:semantic}

Even after structural compression, each column still carries verbose semantic context: a natural-language description, a type, and sample values. In real databases these descriptions are often highly redundant across columns---``timestamp of the last update'', ``timestamp of the most recent update'', ``last-update epoch milliseconds'' are essentially the same tag. The semantic layer of \DbCC{} compresses this redundancy in two stages.

\subsection{Hierarchical keyword extraction}
\label{sec:sem-kw}

For each column $c_i$ with description $d_i$, type $t_i$ and sample values $v_i$, \DbCC{} produces a small set of normalized keywords
\[
K_i = \mathrm{Norm}\!\bigl( K_{\text{desc}}(d_i, v_i) \cup K_{\text{type}}(t_i) \bigr).
\]
The keyword set is intended to act as a \emph{semantic tag} rather than a paraphrase: it should support disambiguation in schema linking, not reproduce the natural-language description.

To preserve cross-table consistency, \DbCC{} extracts keywords \emph{hierarchically}: at the database level when the input fits, then falling back to table level, then to chunk level, with explicit propagation of vocabulary between levels. Database-level extraction is essential: it ensures that synonymous concepts in different tables receive the same tag, which is the single largest source of subsequent component reuse. Normalized type tags such as \texttt{type:variant} and \texttt{type:struct} are exposed explicitly, so that downstream models can reason about composite columns without parsing nested type expressions.

\subsection{Shared semantic-tag componentization}
\label{sec:sem-comp}

After hierarchical extraction, small subsets of keywords are observed to appear across many columns. The semantic-component operator applies \Sgcf{} (Eq.~\eqref{eq:gain}) to the keyword layer, with carriers $\mathcal{O}\!=\!\{c_i\}$ and information units drawn from the keyword vocabulary $\mathcal{V}_M$, so $s(X)$ here counts the number of columns whose keyword set contains $X$. \DbCC{} accepts $X$ as a semantic component iff $s(X)\!\ge\!\tau_s$, $w(X)\!\ge\!\tau_w$ and $\mathrm{gain}(X)\!\ge\!\tau_g$. Each column $c_i$, originally tagged with $K_i$, is rewritten as a pair
\[
\widehat{K}_i = (R_i,\, U_i),\qquad R_i \subseteq \mathcal{C},\ \ U_i = K_i \setminus \bigcup_{X\in R_i} X,
\]
where $R_i$ is the set of references from $c_i$ to accepted components and $U_i$ is the residual keyword set carried inline by $c_i$.

\paragraph*{Frequency-priority component selection} Unlike structural columns, single-column keyword sets are short, so candidate components $X\!\subseteq\!K_i$ can be enumerated directly up to a width cap $w_{\max}$. \DbCC{} then selects components by a \emph{frequency-priority greedy} sweep: sort all admissible candidates by support $s(X)$ (with $\mathrm{gain}(X)$ and $w(X)$ as tie-breakers), and accept each candidate in turn iff it (a) still satisfies the support/width/gain thresholds against the current state and (b) is disjoint on its keyword tokens from every previously accepted component. After each acceptance the residual keyword sets are refreshed and the remaining candidates are re-evaluated. The intuition is that a high-support component produces a thicker shared backbone for the components that follow, so paying its definition cost first amortizes best across the database; the loop terminates after at most $O(|\mathcal{V}|)$ acceptances. Algorithm~\ref{alg:freq} states the procedure formally.

\begin{algorithm}[t]
\caption{Frequency-priority semantic-component selection}
\label{alg:freq}
\begin{algorithmic}[1]
\State \textbf{Input:} columns $\{c_i\}$ with keyword sets $\{K_i\}$; thresholds $(\tau_s,\tau_w,\tau_g)$.
\State Enumerate candidate components $\mathcal{X}\!\leftarrow\!\{X\subseteq K_i: |X|\!\le\!w_{\max}\}$.
\State For each $X\!\in\!\mathcal{X}$, compute $s(X),w(X),\mathrm{gain}(X)$ from~\eqref{eq:gain}.
\State Sort $\mathcal{X}$ by $\bigl(s(X),\mathrm{gain}(X),w(X)\bigr)$ in decreasing order.
\State $\mathcal{C}\!\leftarrow\!\emptyset$.
\For{$X\in\mathcal{X}$ in sorted order}
  \If{$s(X)\!\ge\!\tau_s$ \textbf{and} $w(X)\!\ge\!\tau_w$ \textbf{and} $\mathrm{gain}(X)\!\ge\!\tau_g$ \textbf{and} $X$ disjoint from every $Y\!\in\!\mathcal{C}$}
    \State $\mathcal{C}\!\leftarrow\!\mathcal{C}\cup\{X\}$; remove $X$ from each affected $K_i$.
    \State Recompute $s(\cdot),w(\cdot),\mathrm{gain}(\cdot)$ for the candidates that overlap any modified $K_i$.
  \EndIf
\EndFor
\State \textbf{Output:} component set $\mathcal{C}$; per-column residuals $U_i\!\leftarrow\!K_i\setminus\bigcup_{X\in\mathcal{C}}X$.
\end{algorithmic}
\end{algorithm}

\paragraph*{Why semantic compression cannot be replaced by structural compression} The two layers select components from disjoint vocabularies: structural components come from $\mathcal{V}_S$ (column signatures), semantic components come from $\mathcal{V}_M$ (keyword tokens). A column group such as \texttt{(created\_at, updated\_at, status)} is a structural component; a tag such as \texttt{\{datetime, audit, mutation\}} attached to the column \texttt{updated\_at} is a semantic component. They co-exist in the compressed view, and the ablation in Sec.~\ref{sec:exp-ablation} confirms additive utility.

\section{External-Knowledge Compression}
\label{sec:external}

External documents---data dictionaries, metric glossaries, domain notes---contain information that is essential for resolving ambiguous columns and complex business rules, but per question only a small fraction of any given document is relevant.

\subsection{Question-relevant evidence purification}

\DbCC{} treats the external-knowledge layer as the only \emph{query-aware} component. Given a question $q$ and a document $d$, \DbCC{} extracts evidence
\[
E'(q) = \mathrm{Purify}(q, d),
\]
designed to support schema linking and SQL generation rather than to summarize the document as a whole.

The purification procedure is a thin LLM-driven step that performs typed extraction on $(q,d)$ followed by a deterministic schema-anchoring pass; the full prompt template, the JSON schema each atom must satisfy, and the rule-based fallback that triggers when the LLM emits a malformed atom are all released with our code (see the URL in the abstract). Each piece of evidence it returns is a \emph{typed atom} bound to a specific decision the downstream system has to make when generating SQL. We use four atom types, which together form a minimal cover of where external documents intervene in SQL generation---\emph{which} columns/tables are used, \emph{how} they are combined, \emph{which} values they are filtered by, and \emph{what} question-side vocabulary even refers to:
\begin{itemize}
\item \textbf{Concept-to-element mappings.} A natural-language concept in $q$ is bound to a concrete schema element, driving schema linking. E.g.\ ``net revenue'' $\to$ \texttt{REVENUE.NET\_AMT}.
\item \textbf{Metric definitions.} A business metric named in $q$ is expanded into the expression that computes it, driving aggregation and projection. E.g.\ \emph{MoM growth} $=(\text{cur}-\text{prev})/\text{prev}$.
\item \textbf{Value constraints.} The encoding of a filter value is recovered so the literal in the \texttt{WHERE} clause is correct. E.g.\ \texttt{STATUS} is encoded as \texttt{1=created}, \texttt{2=paid}, \texttt{3=completed}, \texttt{4=refunded}, so the predicate ``completed'' must be written as \texttt{STATUS=3}, not \texttt{STATUS='completed'}.
\item \textbf{Domain definitions.} A term in $q$ that has no direct schema counterpart is expanded into a set of schema-visible elements. E.g.\ ``EU'' $\to$ \{\texttt{DE,\,FR,\,IT,\,ES,\,NL,\dots}\} over \texttt{COUNTRY\_CD}.
\end{itemize}

\paragraph*{Why purification, not retrieval} A retrieval-only baseline returns text passages, which still need to be parsed in context and consume non-trivial tokens. Purification produces \emph{typed evidence atoms} that are easier for downstream linking to consume. This is also the layer at which \Sgcf{} degenerates: the carriers (document chunks) are query-specific, so support is computed over the set of question-relevant chunks rather than all documents. Purification can be viewed as a Text-to-SQL-specific instantiation of retrieval-augmented generation, with the crucial difference that purified targets are \emph{schema-aware}---each evidence atom is anchored to a database element, making it directly usable by schema linking. \DbCC{} can also be combined with arbitrary retrievers (e.g., BM25~\cite{bm25} or dense retrievers~\cite{sbert}) as a pre- or post-processing step.

\section{Integration with Text-to-SQL Pipelines}
\label{sec:integration}

Because \DbCC{} rewrites only the database side, it integrates with a wide spectrum of downstream systems through a single contract: replace the system's view of $D$ with the compressed view $D'$.

\subsection{Insertion points}

Within a typical Text-to-SQL pipeline \DbCC{} can be inserted at two points, mirroring the two stages of the pipeline itself---schema linking and SQL generation. The two insertion points address \emph{different} bottlenecks of the downstream system, and \DbCC{} contributes a different aspect of $D'$ to each.

\begin{itemize}
\item \textbf{Pre-linking (\emph{shrinks the candidate space \emph{and} sharpens token-level alignment}).} The system must decide which database elements are relevant to the question. We consider schema linking broadly, covering both granularities: token-level alignment over the full schema~\cite{ratsql,resdsql,bridge,rslsql} and retrieval-style subsetting that narrows a far larger candidate space~\cite{crush4sql,linkalign,codes,dbexplore}. \DbCC{} contributes to both. At the size axis, replacing $S$ with the structurally compressed $S'$ collapses hundreds of sharded variants into one canonical instance and exposes each repeated audit/status block once instead of dozens of times; the linker's input is smaller, and its recall ceiling rises because near-duplicate candidates that previously competed for the top-$k$ slots are now merged. At the density axis, hierarchical semantic tags in $M'$ replace verbose, inconsistently worded column descriptions, and purified evidence atoms in $E'$ provide explicit NL$\to$element anchors; the linker sees fewer, more discriminative cues per candidate.
\item \textbf{Pre-generation (\emph{reduces prompt cost and agent loops}).} If the system directly prompts an LLM~\cite{c3sql,dinsql,actsql,dailsql,petsql,chasesql} or runs an agentic flow~\cite{chess,reforce,macsql,autolink}, $D'$ shortens the prompt, focuses attention on discriminative information, and---for agents---cuts the number of database-exploration steps needed before the model commits to a plan.
\end{itemize}

The two insertion points compose: a system that does linking and LLM generation in series can ingest $D'(q)$ once and benefit at both stages, because $S',M',E'(q)$ are produced as a single self-contained view.

\subsection{Cost model and minimal assumptions}

\DbCC{} does not modify the parser, decoder, schema-linking module, agent controller or post-processor of the downstream system: the rewrite is reversible (any system can expand $S'$ back to $S$), no additional supervision is needed (structural/semantic compression is unsupervised, evidence purification needs only the question), and no per-query online learning state is kept on the database side beyond the cached Phase~I view. Let $c_{\mathrm{I}}(\mathcal{D})$ be the offline cost of Phase~I, $c_{\mathrm{II}}$ the per-query cost of Phase~II, $\Delta c$ the per-query token saving, and $N$ the expected number of queries against $\mathcal{D}$. \DbCC{} is profitable iff $N \cdot \Delta c \ge c_{\mathrm{I}}(\mathcal{D}) + N\cdot c_{\mathrm{II}}$, so the break-even number of queries is
\[
N^{\ast} \;=\; \bigl\lceil c_{\mathrm{I}}(\mathcal{D})\,/\,(\Delta c - c_{\mathrm{II}}) \bigr\rceil.
\]
$N^{\ast}$ is small whenever $\Delta c \gg c_{\mathrm{II}}$, which is the regime our compressors target; we measure $N^{\ast}$ empirically in Sec.~\ref{sec:exp}.

\section{Experiments}
\label{sec:exp}

We design our study around five questions:
(Q1)~\textbf{How much} can \DbCC{} compress real database context, and \textbf{at what utility cost or gain} for downstream schema linking?
(Q2)~\textbf{Are the three operators truly orthogonal}, and how do they compose?
(Q3)~\textbf{Does \DbCC{} transfer} across different LLMs, different benchmarks, and different downstream pipelines, and is the offline phase \textbf{economically amortizable}?
(Q4)~\textbf{What does the cost--utility trade-off look like}---is there a sweet spot, and is the curve monotone?
(Q5)~\textbf{Where does \DbCC{} still fail}, and which residual errors should the next iteration target?

\subsection{Setup}
\label{sec:exp-setup}

\paragraph*{Benchmarks} We use \textbf{Spider~2.0-Snow}~\cite{spider2} as our primary benchmark. It is a real-workflow benchmark on Snowflake with 152 enterprise databases, 547 questions and an average of 812.1~columns per database; the average gold SQL involves 161.8 tokens and 6.8 function calls. The column-count distribution is heavy-tailed: while the mean is dominated by a long tail of small databases, a substantial minority sits in the $\ge\!10^4$ regime that motivates this work. Following Sec.~\ref{sec:intro-largeness}, we partition the databases into three buckets by column count: \textsc{Small} ($<\!1{,}000$ columns; $76$ DBs), \textsc{Medium} ($1{,}000$--$10{,}000$ columns; $53$ DBs) and \textsc{Large} ($\ge\!10{,}000$ columns; $23$ DBs). For cross-benchmark generalization we use the full \textbf{BIRD}~\cite{bird} benchmark---all of its 95 databases and 1{,}534 development questions, with no column-count filtering.

\paragraph*{Models} We use \textbf{DeepSeek-V3.2} as the primary LLM for both Phase~I (semantic keyword extraction, evidence purification) and the downstream task (schema linking, SQL generation). We additionally evaluate cross-LLM robustness with \textbf{GPT-4o} and \textbf{Claude-Opus-4.7}.

\paragraph*{Tasks and metrics} The primary task is schema linking: given a database context and a question, the LLM must output the set of schema elements (tables and columns) needed to answer the question. All metrics below follow the conventions established in recent schema-linking work---in particular RSL-SQL~\cite{rslsql} and Apex-SQL~\cite{apexsql}---so that our numbers are directly comparable to theirs. We report:
\begin{itemize}
\item \textbf{Strict Recall Rate (SRR)}: per-instance indicator of whether the prediction covers \emph{all} gold schema elements.
\item \textbf{Non-Strict Recall (NSR)} and \textbf{Non-Strict Precision (NSP)}: micro-level recall and precision over predicted vs.\ gold elements.
\item \textbf{Token cost (Tok)}: average input tokens per request; reported in K (thousand) or M (million).
\item \textbf{Execution Accuracy (EX)}: for end-to-end Text-to-SQL experiments, the fraction of generated SQL queries whose execution result matches the gold query.
\end{itemize}

\paragraph*{Default \DbCC{} configuration} Structural compression uses \emph{template hierarchy}, semantic compression uses the frequency-priority component selection of Algorithm~\ref{alg:freq} with thresholds $(\tau_s,\tau_w,\tau_g)=(2,2,1)$, and external-knowledge compression uses the question-driven evidence purification described in Sec.~\ref{sec:external}.

\subsection{Structural Compression}
\label{sec:exp-struct}

We first evaluate the two structural operators in isolation. Table~\ref{tab:struct} reports schema-linking utility and end-to-end EX, against the uncompressed schema, on the three database buckets. The two informative axes are the input cost (Tok) and the resulting utility (SRR/NSR/NSP/EX); their non-monotone trade-off is examined in Sec.~\ref{sec:exp-pareto}.

The two operators capture complementary redundancy patterns and select from disjoint candidate spaces (single tables for factorization vs.\ table clusters for hierarchy). Component factorization is stronger on \textsc{Small} and \textsc{Large} databases (where audit/status groups dominate), template hierarchy on \textsc{Medium} (where sharded families dominate). On \textsc{Small} the gain is essentially in cost: tokens drop by $\sim\!22\%$ while SRR/NSR/NSP stay within $1$--$3\%$ of the baseline. On \textsc{Medium}, template hierarchy already dominates: tokens drop by \Up{82.0\%} ($166.0$K~$\to$~$29.9$K) and SRR rises by \Up{5.2\%}, NSR by \Up{14.1\%}. On \textsc{Large} the contrast is striking: the uncompressed schema does not fit and yields SRR$\,=\,$0.0\%, while template hierarchy reaches $47.8\%$ SRR with a \Up{96.1\%} token reduction---compression is a \emph{prerequisite} for the model to attempt the task at all.

\begin{table*}[t]
\centering
\caption{Structural compression on Spider~2.0-Snow: schema-linking utility and end-to-end EX, vs.\ the uncompressed schema. LLM: DeepSeek-V3.2. SRR/NSR/NSP/EX in \%; tokens in K or M. Best per metric per bucket in \textbf{bold}, second-best \underline{underlined}.}
\label{tab:struct}
\footnotesize
\setlength{\tabcolsep}{3pt}
\begin{tabular}{@{}lrrrrrrrrrrrrrrr@{}}
\toprule
& \multicolumn{5}{c}{\textsc{Small}} & \multicolumn{5}{c}{\textsc{Medium}} & \multicolumn{5}{c}{\textsc{Large}}\\
\cmidrule(lr){2-6}\cmidrule(lr){7-11}\cmidrule(lr){12-16}
Method & SRR & NSR & NSP & Tok & EX & SRR & NSR & NSP & Tok & EX & SRR & NSR & NSP & Tok & EX\\
\midrule
Raw schema      & \textbf{61.1}    & \underline{84.3} & \textbf{78.6}    & 9.6K             & 28.6             & 44.8             & 67.3          & 57.1          & 166.0K          & 9.4              & 0.0           & 2.9           & 50.0          & 2.6M             & 0.0\\
Comp.\ factor.\ & 58.3             & 83.0             & \underline{78.3} & \textbf{7.4K}    & \underline{29.5} & \underline{49.0} & 78.0          & 48.4          & \underline{33.8K}& \underline{15.8} & 43.5          & 67.6          & 63.7          & \underline{82.9K}& \underline{10.7}\\
Templ.\ hier.   & \underline{60.6} & 83.8             & 75.5             & \underline{7.5K} & \textbf{30.4}    & \textbf{50.0}    & \textbf{81.4} & \textbf{57.4} & \textbf{29.9K}  & \textbf{17.2}    & \textbf{47.8} & \textbf{69.5} & \textbf{71.6} & 100.7K           & \textbf{11.6}\\
\bottomrule
\end{tabular}
\end{table*}

\paragraph*{Per-database-type analysis}
Classifying the 152 databases by structural fingerprint into \textsc{Wide-Heavy} (max column-signature support $s_{\max}\!\ge\!10$ and table-cluster ratio $<\!0.3$; 58 DBs), \textsc{Partition-Heavy} (table-cluster ratio $\ge\!0.3$; 41 DBs) and \textsc{Mixed} (the remaining 53 DBs) confirms what the design predicts. Factorization wins on \textsc{Wide-Heavy} (SRR $52.1$ vs.\ $49.0$ for hierarchy; EX $15.4$ vs.\ $13.7$); hierarchy wins on \textsc{Partition-Heavy} (SRR $53.6$ vs.\ $46.2$; EX $16.8$ vs.\ $11.5$); on \textsc{Mixed} neither subsumes the other and applying both is preferable (SRR $51.9$, EX $15.1$). Hence the two operators select components from disjoint candidate spaces and are jointly needed.

\subsection{Semantic Compression}
\label{sec:exp-sem}

Table~\ref{tab:sem} reports schema-linking utility and end-to-end EX of semantic compression vs.\ the raw representation, on the same three buckets. Tokens drop by $46.5\%$ on \textsc{Small} ($9.6$K~$\to$~$5.1$K) and by roughly $73\%$ on both \textsc{Medium} and \textsc{Large} ($166.0$K~$\to$~$43.6$K, $2.6$M~$\to$~$712.4$K), confirming that semantic redundancy scales with the number of columns rather than with the size of any individual description. Utility stays close to the raw baseline on \textsc{Small} (where it is already saturated) and \emph{rises} on the larger buckets (SRR $44.8\!\to\!49.0$ on \textsc{Medium}; $0.0\!\to\!8.7$ on \textsc{Large}). Semantic-only compression is not sufficient to make \textsc{Large} databases tractable in isolation; the full benefit emerges only when combined with structural compression (Sec.~\ref{sec:exp-ablation}).

\begin{table*}[t]
\centering
\caption{Semantic compression on Spider~2.0-Snow: schema-linking utility and end-to-end EX, vs.\ the raw semantic representation. LLM: DeepSeek-V3.2. SRR/NSR/NSP/EX in \%; tokens in K or M. Best per metric per bucket in \textbf{bold}.}
\label{tab:sem}
\footnotesize
\setlength{\tabcolsep}{3pt}
\begin{tabular}{@{}lrrrrrrrrrrrrrrr@{}}
\toprule
& \multicolumn{5}{c}{\textsc{Small}} & \multicolumn{5}{c}{\textsc{Medium}} & \multicolumn{5}{c}{\textsc{Large}}\\
\cmidrule(lr){2-6}\cmidrule(lr){7-11}\cmidrule(lr){12-16}
Method & SRR & NSR & NSP & Tok & EX & SRR & NSR & NSP & Tok & EX & SRR & NSR & NSP & Tok & EX\\
\midrule
Raw semantic   & \textbf{61.1} & 84.3          & \textbf{78.6} & 9.6K          & 28.6          & 44.8          & 67.3          & \textbf{57.1} & 166.0K         & 9.4           & 0.0          & 2.9           & 50.0          & 2.6M             & 0.0\\
Keyword comp.\ & 60.6          & \textbf{85.0} & 74.8          & \textbf{5.1K} & \textbf{29.1} & \textbf{49.0} & \textbf{76.4} & 47.5          & \textbf{43.6K} & \textbf{14.2} & \textbf{8.7} & \textbf{13.3} & \textbf{60.9} & \textbf{712.4K} & \textbf{2.6}\\
\bottomrule
\end{tabular}
\end{table*}


\subsection{External-Knowledge Compression}
\label{sec:exp-ek}

We evaluate evidence purification on the subset of Spider~2.0-Snow questions that reference an external document. Table~\ref{tab:ek} reports the resulting schema-linking utility and end-to-end EX on the same subset.

At the \emph{prompt} level the token savings from purification are essentially negligible---$229.2$K~$\to$~$228.1$K input tokens---because the external document is only one of several prompt components and the rest (schema, semantic descriptors, in-context examples) dominates. Despite this, replacing raw documents with purified evidence improves SRR by \Up{2.7\%}, NSR by \Up{14.7\%} and NSP by \Up{7.2\%} on the documented subset, with a corresponding gain of \Up{2.6\%} in end-to-end EX. This is the qualitative claim of Sec.~\ref{sec:external} made empirical: the value of external-knowledge compression is not in length reduction but in the \emph{focus shift}---business rules, concept-to-element mappings and value constraints become explicit, schema-anchored evidence atoms rather than being scattered across long prose.

\begin{table}[t]
\centering
\caption{External-knowledge compression on the Spider~2.0-Snow subset with external knowledge. LLM: DeepSeek-V3.2. ``Tok'' is the prompt-level input-token count. Best per metric in \textbf{bold}.}
\label{tab:ek}
\small
\setlength{\tabcolsep}{6pt}
\begin{tabular}{@{}lrrrrr@{}}
\toprule
Method & SRR & NSR & NSP & Tok & EX\\
\midrule
Raw documents     & 49.3          & 73.9          & 69.4          & 229.2K          & 12.7\\
Purified evidence & \textbf{52.0} & \textbf{88.6} & \textbf{76.6} & \textbf{228.1K} & \textbf{15.3}\\
\bottomrule
\end{tabular}
\end{table}

\subsection{Ablation and Orthogonality}
\label{sec:exp-ablation}

We now evaluate the three operators jointly with a clean $2^3$-style ablation, where each operator is independently turned on or off. Structural compression uses template hierarchy, semantic uses frequency-priority, and external knowledge uses purified evidence; \emph{nothing else changes between rows}, so the differences are attributable to the operators themselves. Table~\ref{tab:ablation} summarizes all eight configurations on the three buckets.

\begin{table*}[t]
\centering
\caption{Fine-grained ablation of \DbCC{} on Spider~2.0-Snow. LLM: DeepSeek-V3.2. Columns S/Sem/EK indicate whether structural, semantic and external-knowledge operators are on. The row \texttt{(\ding{55},\ding{55},\ding{55})} is the uncompressed baseline; \texttt{(\ding{51},\ding{51},\ding{51})} is full \DbCC{}. Best per metric per bucket in \textbf{bold}, second-best \underline{underlined}.}
\label{tab:ablation}
\scriptsize
\setlength{\tabcolsep}{2.5pt}
\begin{tabular}{@{}cccrrrrrrrrrrrrrrr@{}}
\toprule
& & & \multicolumn{5}{c}{\textsc{Small}} & \multicolumn{5}{c}{\textsc{Medium}} & \multicolumn{5}{c}{\textsc{Large}}\\
\cmidrule(lr){4-8}\cmidrule(lr){9-13}\cmidrule(lr){14-18}
S & Sem & EK & SRR & NSR & NSP & Tok & EX & SRR & NSR & NSP & Tok & EX & SRR & NSR & NSP & Tok & EX\\
\midrule
\ding{55} & \ding{55} & \ding{55} & \textbf{61.1} & \underline{84.3} & \textbf{78.6} & 9.6K  & 28.6 & 44.8 & 67.3 & 57.1 & 166.0K & 9.4 & 0.0  & 2.9  & 50.0 & 2.6M & 0.0 \\
\ding{55} & \ding{55} & \ding{51} & \textbf{62.3} & \underline{84.3} & 76.1 & 9.4K  & 29.0 & 45.8 & 77.8 & \textbf{62.2} & 165.4K & 9.9 & 4.3  & 4.3  & 60.0 & 2.6M & 1.0 \\
\ding{55} & \ding{51} & \ding{55} & 60.6 & \textbf{85.0} & 74.8 & 5.1K  & 29.1 & 49.0 & 76.4 & 47.5 & 43.6K  & 14.2 & 8.7  & 13.3 & 60.9 & 712.4K & 2.6 \\
\ding{51} & \ding{55} & \ding{55} & 60.6 & 83.8 & 75.5 & 7.5K  & 30.4 & 50.0 & 81.4 & 57.4 & 29.9K  & \textbf{17.2} & 47.8 & 69.5 & 71.6 & 100.7K & \textbf{11.6} \\
\ding{55} & \ding{51} & \ding{51} & 60.0 & 82.0 & 75.3 & 5.0K  & 29.3 & 47.9 & 74.2 & 48.7 & 43.1K  & 12.7 & 13.0 & 24.3 & 71.8 & 716.3K & 3.4 \\
\ding{51} & \ding{55} & \ding{51} & 59.1 & 83.0 & 74.7 & 7.4K  & 30.3 & 51.0 & 78.9 & 54.6 & 29.3K  & 14.6 & 43.5 & 75.7 & 69.1 & 100.6K & 9.4 \\
\ding{51} & \ding{51} & \ding{55} & \textbf{61.1} & 83.3 & 71.4 & 4.7K  & \underline{30.7} & \underline{55.2} & 77.3 & 58.1 & 14.9K  & \underline{16.4} & \underline{52.2} & \underline{76.7} & \underline{72.9} & 34.9K & \underline{11.0} \\
\ding{51} & \ding{51} & \ding{51} & \underline{61.7} & 82.4 & 71.8 & \textbf{4.5K}  & \textbf{31.2} & \textbf{56.3} & \textbf{85.9} & 55.7 & \textbf{14.3K}  & \textbf{17.2} & \textbf{56.5} & \textbf{78.6} & \textbf{73.3} & \textbf{34.7K} & \textbf{11.6} \\
\bottomrule
\end{tabular}
\end{table*}

\paragraph*{Cumulative gains}
On \textsc{Medium} databases, full \DbCC{} improves SRR by \Up{11.5\%} and NSR by \Up{18.6\%} over the uncompressed baseline, while reducing tokens by \Up{91.4\%} ($166.0$K~$\to$~$14.3$K). On \textsc{Large} databases the SRR gap widens to \Up{56.5\%}---essentially the difference between \emph{not running at all} and a working linker---together with a \Up{98.7\%} token reduction. On \textsc{Small} databases the utility is already saturated by the baseline; here \DbCC{} preserves SRR/NSR within $1\%$--$2\%$ while still cutting tokens by \Up{53.1\%}, indicating that the operators are not actively harmful on inputs that did not need compression.

\paragraph*{Orthogonality}
We measure orthogonality via the index $\mathrm{Orth}(A,B) = \bigl(\Delta(A\!\cup\!B) - \max(\Delta A,\Delta B)\bigr)/\min(\Delta A,\Delta B)$ in terms of SRR gain over the baseline. Computing on the \textsc{Medium} bucket using Table~\ref{tab:ablation}:
\begin{itemize}
\item $\Delta(\text{S})\!=\!5.2$, $\Delta(\text{Sem})\!=\!4.2$, $\Delta(\text{EK})\!=\!1.0$.
\item $\mathrm{Orth}(\text{S},\text{Sem})\!=\!(10.4\!-\!5.2)/4.2 \!=\! 1.24$,
      $\mathrm{Orth}(\text{S},\text{EK})\!=\!(6.2\!-\!5.2)/1.0\!=\!1.00$,
      $\mathrm{Orth}(\text{Sem},\text{EK})\!=\!(3.1\!-\!4.2)/1.0\!=\!-1.10$.
\end{itemize}
Two pairs are clearly additive (Orth $>0$) and one pair (semantic+EK without structural) is mildly antagonistic, which is interpretable: both semantic and EK operators rely on a high-quality column-level alignment, which only structural compression can deliver on \textsc{Large} schemas. This interpretation is consistent with the \textsc{Large} row of Table~\ref{tab:ablation}, where structural compression is what unlocks SRR above $40\%$.

\subsection{End-to-End Integration}
\label{sec:exp-integration}

To check whether a database-side rewrite translates to end-to-end gains, we plug \DbCC{} into three recent Text-to-SQL systems---\textbf{AutoLink}~\cite{autolink}, \textbf{ReFoRCE}~\cite{reforce} and \textbf{Apex-SQL}~\cite{apexsql}---without modifying their pipelines, decoders, or post-processing. We restrict to Spider~2.0-Snow databases with $>$5{,}000 columns, where context organization most aggressively dominates accuracy.

\begin{table}[t]
\centering
\caption{End-to-end plug-in integration with three recent Text-to-SQL systems on Spider~2.0-Snow databases with $>$5{,}000 columns. LLM: DeepSeek-V3.2. Best within each system in \textbf{bold}.}
\label{tab:e2e}
\small
\setlength{\tabcolsep}{6pt}
\begin{tabular}{@{}llrr@{}}
\toprule
System    & DB context        & EX     & Tok\\
\midrule
AutoLink~\cite{autolink}  & Raw       & 19.6\%          & 61.0K\\
AutoLink                  & + \DbCC{} & \textbf{21.4\%} & \textbf{26.2K}\\
ReFoRCE~\cite{reforce}    & Raw       & 16.1\%          & 16.6K\\
ReFoRCE                   & + \DbCC{} & \textbf{17.9\%} & \textbf{4.4K}\\
Apex-SQL~\cite{apexsql}   & Raw       & 43.7\%          & 48.3K\\
Apex-SQL                  & + \DbCC{} & \textbf{45.6\%} & \textbf{19.5K}\\
\bottomrule
\end{tabular}
\end{table}

Table~\ref{tab:e2e} shows that on all three systems, simply replacing the raw context with the \DbCC{}-compressed view improves EX (by \Up{1.8\%}, \Up{1.8\%} and \Up{1.9\%} respectively) while substantially reducing token consumption (\Up{57.0\%} for AutoLink, \Up{73.5\%} for ReFoRCE and \Up{59.6\%} for Apex-SQL). Apex-SQL is by a wide margin the strongest baseline on raw inputs ($43.7\%$ EX, more than twice that of AutoLink), reflecting the effectiveness of its agentic exploration loop at narrowing the database context per turn; even at this much higher operating point, \DbCC{} compounds with that loop and adds a measurable EX gain at roughly $40\%$ of the token cost. The absolute EX gains are modest ($\le\!2\%$) and, as expected, smaller than the schema-linking improvements of Sec.~\ref{sec:exp-ablation}: end-to-end EX is influenced by SQL planning, dialect handling, error recovery and post-processing, all of which are downstream of database context. \DbCC{} only changes the input these systems consume, yet the EX improvement is positive on every one of the three systems, suggesting that real Text-to-SQL stacks are bottlenecked by context quality even when they invest in agentic flows or self-refinement.

\subsection{Cross-LLM Robustness}
\label{sec:exp-llm}

We evaluate \DbCC{} under two additional commercial LLMs, \textbf{GPT-4o} and \textbf{Claude-Opus-4.7}, on the same Spider~2.0-Snow setting (Table~\ref{tab:llm}). Two patterns emerge across both models. (i)~On \textsc{Small} databases, where the uncompressed prompt already fits and is comprehensible, the relative gain shrinks because larger models already handle redundancy gracefully. (ii)~On \textsc{Medium} and especially \textsc{Large} databases, where the issue is whether the prompt fits at all, the gap widens---a $2.6$M-token raw prompt remains beyond the input budget of every LLM we tested, so absolute SRR stays at $0$ without compression and rises sharply once \DbCC{} is plugged in. Claude-Opus-4.7 is the strongest of the three on raw inputs and remains the strongest under \DbCC{}, but the relative improvement that \DbCC{} contributes is essentially constant across LLMs, supporting our framing: \DbCC{} addresses an input-side problem and its benefit is largely independent of downstream model capacity.

\begin{table*}[t]
\centering
\caption{Cross-LLM robustness of \DbCC{} on Spider~2.0-Snow. Within each LLM, best per metric per bucket in \textbf{bold} (higher is better for SRR/EX, lower for Tok).}
\label{tab:llm}
\small
\setlength{\tabcolsep}{4pt}
\begin{tabular}{@{}llrrrrrrrrr@{}}
\toprule
LLM & Method & \multicolumn{3}{c}{\textsc{Small}} & \multicolumn{3}{c}{\textsc{Medium}} & \multicolumn{3}{c}{\textsc{Large}}\\
\cmidrule(lr){3-5}\cmidrule(lr){6-8}\cmidrule(lr){9-11}
& & SRR & Tok & EX & SRR & Tok & EX & SRR & Tok & EX\\
\midrule
DeepSeek-V3.2           & Raw       & 61.1          & 9.6K          & 28.6          & 44.8          & 166K           & 9.4           & 0.0           & 2.6M            & 0.0\\
DeepSeek-V3.2           & + \DbCC{} & \textbf{61.7} & \textbf{4.5K} & \textbf{31.2} & \textbf{56.3} & \textbf{14.3K} & \textbf{17.2} & \textbf{56.5} & \textbf{34.7K}  & \textbf{11.6}\\
GPT-4o        & Raw       & 65.4          & 9.6K          & 32.1          & 49.6          & 166K           & 11.8          & 0.0           & 2.6M            & 0.0\\
GPT-4o        & + \DbCC{} & \textbf{65.8} & \textbf{4.5K} & \textbf{33.5} & \textbf{60.1} & \textbf{14.3K} & \textbf{19.6} & \textbf{60.7} & \textbf{34.7K}  & \textbf{13.4}\\
Claude-Opus-4.7 & Raw       & 67.2          & 9.6K          & 33.8          & 51.4          & 166K           & 12.6          & 0.0           & 2.6M            & 0.0\\
Claude-Opus-4.7 & + \DbCC{} & \textbf{67.6} & \textbf{4.5K} & \textbf{35.1} & \textbf{62.3} & \textbf{14.3K} & \textbf{20.7} & \textbf{63.1} & \textbf{34.7K}  & \textbf{14.5}\\
\bottomrule
\end{tabular}
\end{table*}

\subsection{Cross-Benchmark Generalization}
\label{sec:exp-bench}

We evaluate \DbCC{} on the full \textbf{BIRD} benchmark (Table~\ref{tab:bench}), with no column-count filtering. We use GPT-4o as the LLM here (rather than DeepSeek-V3.2 as in the rest of the experiments) because GPT-4o is the model on which existing BIRD numbers are most commonly reported, making the \emph{Raw} row directly comparable to the literature; the cross-LLM evidence in Sec.~\ref{sec:exp-llm} confirms that the $\Delta$-pattern is essentially LLM-agnostic. Although BIRD has shorter average column counts than Spider~2.0-Snow, it features richer business semantics and external knowledge, so we expect proportionally larger gains from semantic and EK compression and slightly smaller gains from structural compression.

\begin{table}[t]
\centering
\caption{Cross-benchmark generalization of \DbCC{} on Spider~2.0-Snow and the full BIRD development set. LLM: GPT-4o. Within each dataset, best per metric in \textbf{bold} (higher is better for SRR/NSR/NSP/EX, lower for Tok).}
\label{tab:bench}
\small
\setlength{\tabcolsep}{3pt}
\begin{tabular}{@{}llrrrrr@{}}
\toprule
Dataset & Method & SRR & NSR & NSP & Tok & EX\\
\midrule
Spider~2.0-Snow & Raw       & 38.2 & 56.6 & 55.7 & 925K & 14.6\\
Spider~2.0-Snow & + \DbCC{} & \textbf{60.9} & \textbf{81.0} & \textbf{67.4} & \textbf{18.0K} & \textbf{22.3}\\
BIRD  & Raw       & 53.2 & 70.4 & 60.8 & 38.6K & 53.6\\
BIRD  & + \DbCC{} & \textbf{60.5} & \textbf{77.1} & \textbf{63.9} & \textbf{12.4K} & \textbf{58.2}\\
\bottomrule
\end{tabular}
\end{table}

The qualitative pattern transfers: \DbCC{} consistently improves all three utility metrics while reducing tokens. The relative SRR gain is smaller on BIRD (\Up{7.3\%} vs.\ \Up{22.7\%} on Spider~2.0-Snow) because BIRD's databases---most of which fall well below the column-count regime that motivates this work---leave less raw structural redundancy to exploit; the gain on BIRD is therefore driven mostly by the semantic and external-knowledge operators rather than by structural rewriting.

\subsection{Comparison with Schema Linking and Pruning Baselines}
\label{sec:exp-baselines}

We compare \DbCC{} against three representative query-aware baselines: \textbf{Crush4SQL}~\cite{crush4sql}, \textbf{LinkAlign}~\cite{linkalign} and \textbf{RSL-SQL}~\cite{rslsql}. We also report \emph{stacked} settings where \DbCC{}'s rewritten schema is fed into the baseline as input, to test whether the methods are orthogonal in practice.

\begin{table}[t]
\centering
\caption{Comparison with query-aware schema-linking baselines on Spider~2.0-Snow. LLM: GPT-4o. Best per metric in \textbf{bold} (higher is better for SRR/NSR/NSP/EX, lower for Tok).}
\label{tab:baselines}
\small
\setlength{\tabcolsep}{3pt}
\begin{tabular}{@{}lrrrrr@{}}
\toprule
Method                          & SRR & NSR & NSP & Tok & EX\\
\midrule
Raw                             & 38.2 & 56.6 & 55.7 & 925K & 14.6\\
Crush4SQL~\cite{crush4sql}      & 47.5 & 67.4 & 58.1 & 102K & 17.9\\
LinkAlign~\cite{linkalign}      & 51.0 & 71.5 & 60.4 & 86K  & 19.0\\
RSL-SQL~\cite{rslsql}           & 49.5 & 70.1 & 61.7 & 96K  & 18.6\\
\textbf{\DbCC{}}                & 60.9 & 81.0 & 67.4 & \textbf{18.0K} & 22.3\\
\midrule
\DbCC{} + Crush4SQL              & 62.0 & 82.4 & \textbf{68.9} & 19.4K & 22.9\\
\DbCC{} + LinkAlign              & \textbf{63.6} & \textbf{83.5} & 68.5 & \textbf{17.6K} & \textbf{23.6}\\
\bottomrule
\end{tabular}
\end{table}

Three observations follow. (i)~Standalone \DbCC{} already outperforms every query-aware baseline on every metric while using $\sim\!5\!\times$ fewer tokens, supporting our framing that \emph{database-side rewriting tackles a different bottleneck} from query-side filtering. (ii)~Stacking \DbCC{} with a query-aware baseline yields further gains (up to \Up{2.7\%} SRR over standalone \DbCC{}), confirming that the two families of methods are orthogonal: query-agnostic compression provides a denser candidate space, on which query-aware schema linking can then operate more reliably. (iii)~The token-cost gap remains: even the stacked variants stay below 20K tokens, two orders of magnitude smaller than the raw schema.

\subsection{Cost--Utility Pareto Frontier}
\label{sec:exp-pareto}

\begin{figure}[t]
\centering
\includegraphics[width=0.8\columnwidth]{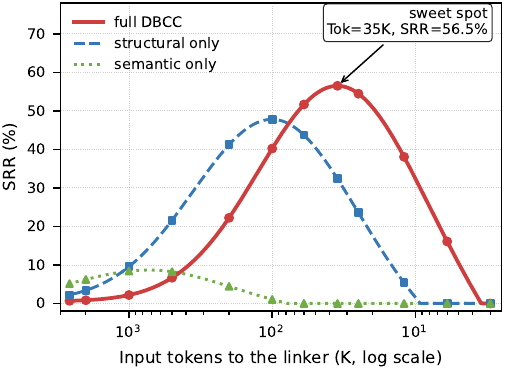}
\caption{Cost--utility Pareto frontier on \textsc{Large} databases of Spider~2.0-Snow. LLM: DeepSeek-V3.2. Three threshold sweeps are shown: structural-only, semantic-only, and full \DbCC{}. The horizontal axis is the input-token cost (log scale, smaller-is-better, hence inverted); the vertical axis is SRR. SRR rises non-monotonically as the prompt shrinks, peaks at Tok$\,\approx\,$34.7K for full \DbCC{}, and drops at extreme over-compression.}
\label{fig:pareto}
\end{figure}

To map out the trade-off, we sweep the thresholds of each operator and plot \emph{input-token cost} vs.\ SRR on the \textsc{Large} bucket. The sampled points (Fig.~\ref{fig:pareto}) reveal a non-monotone curve: shrinking the prompt initially \emph{raises} utility, peaks at Tok$\,\approx\,$34.7K for full \DbCC{} (matching the row in Table~\ref{tab:ablation}), and only at extreme over-compression (Tok\,$\lesssim\!10$K) starts to drop, when discriminative columns are absorbed into too-small components---a manifestation of the \emph{Lost-in-the-Middle} effect~\cite{lostmiddle}.

\subsection{Offline Cost and Amortization}
\label{sec:exp-amort}

We finally check whether the offline cost of Phase~I is justified (Table~\ref{tab:amort}). The Phase~I cost is dominated by LLM-based hierarchical keyword extraction (Sec.~\ref{sec:sem-kw}); the structural rewrite is combinatorial and negligible, and Phase~II adds a small per-question cost.

\begin{table}[t]
\centering
\caption{Offline cost and amortization of \DbCC{} on Spider~2.0-Snow. LLM: DeepSeek-V3.2. $N^\ast$ = break-even number of queries per database. ``P-I~Tok'' and ``P-I~s'' are Phase~I token cost and wall-clock time per database; ``P-II~Tok'' is Phase~II cost per question; ``$\Delta$/q'' is per-query saving on the downstream pipeline.}
\label{tab:amort}
\footnotesize
\setlength{\tabcolsep}{4pt}
\begin{tabular}{@{}lrrrrr@{}}
\toprule
Scale & P-I~Tok & P-I~s & P-II~Tok & $\Delta$/q & $N^\ast$\\
\midrule
\textsc{Small}  & 14K   & 26   & 0.7K & 4.6K   & 4 \\
\textsc{Medium} & 138K  & 240  & 1.3K & 152K   & 1 \\
\textsc{Large}  & 2.0M  & 3,500 & 1.6K & 2.57M  & 1 \\
\bottomrule
\end{tabular}
\end{table}

The break-even $N^{\ast}$ is small: even on \textsc{Small} databases, $\sim$4 questions already amortise the offline cost; on \textsc{Medium}/\textsc{Large} databases $N^{\ast}\!=\!1$, i.e.\ \DbCC{} is net-positive after a single query, since $\Delta c$ scales with database size while $c_{\mathrm{II}}$ does not.

\subsection{Failure-Mode Analysis}
\label{sec:exp-failures}

\begin{figure}[!htbp]
\centering
\includegraphics[width=\columnwidth]{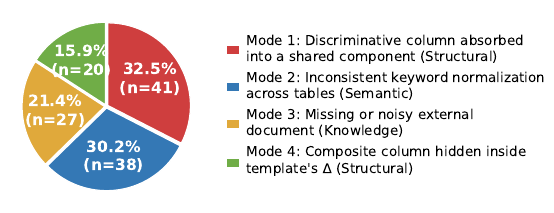}
\caption{Distribution of failure modes of full \DbCC{} on Spider~2.0-Snow. LLM: DeepSeek-V3.2. Slices show the share of each failure mode within the schema-linking error subset; absolute case counts in parentheses.}
\label{fig:failures}
\end{figure}

We finally take stock of where \DbCC{} fails. Inspecting the schema-linking errors of the full \DbCC{} configuration on Spider~2.0-Snow yields four recurring failure modes, whose distribution is shown in Fig.~\ref{fig:failures}.

\textbf{Mode~1} arises when an audit-style column that happens to be discriminative for a given query is folded into a frequent component; the column is still recoverable by expansion, but the compressed view does not surface it as salient. \textbf{Mode~2} reflects the limits of LLM-driven keyword extraction: synonyms can survive normalization at the table level but disagree across tables, hurting cross-table linking. \textbf{Mode~3} is upstream of \DbCC{} and is shared by any external-knowledge-aware system. \textbf{Mode~4} is specific to the template hierarchy operator: when the parent template is large, the per-derived $\Delta$ may include composite columns that are easy to overlook.

\subsection{Summary, limitations and future work}
We summarise the answers to the five questions of Sec.~\ref{sec:exp}. \textbf{(Q1)}~On real \textsc{Large} databases \DbCC{} cuts linker input by $98\%$ and lifts strict recall from $0\%$ to $56.5\%$ (DeepSeek-V3.2)/$63.1\%$ (Claude-Opus-4.7), with end-to-end EX on Spider~2.0-Snow rising from $14.6\%$ to $22.3\%$ under GPT-4o. \textbf{(Q2)}~The three operators are largely orthogonal ($\mathrm{Orth}\!>\!0$ on two of three pairs; layer-specific failure modes). \textbf{(Q3)}~The benefit transfers across LLMs, benchmarks, and three downstream systems, stacks additively with three schema linkers, and Phase~I amortises after one query on medium/large databases. \textbf{(Q4)}~The cost--utility curve is non-monotone with a sweet spot at Tok$\,\approx\,$34.7K. \textbf{(Q5)}~Discriminative-column absorption and cross-table keyword inconsistency drive $62.7\%$ of residual errors, motivating a query-aware extension as future work.

\section{Discussion and Conclusion}
\label{sec:conclusion}

\textbf{Discussion.} \DbCC{} suggests two takeaways for the Text-to-SQL community. \emph{First}, the bottleneck has moved to the database side: even the strongest LLMs cannot ingest a $2.6$M-token schema, the \emph{Lost-in-the-Middle} effect~\cite{lostmiddle} makes verbosity actively hurtful, and the resulting ceiling can be raised \emph{without} touching the model or the linker. \emph{Second}, classical database thinking carries over to the LLM era: our offline/online split mirrors the indexes / materialised-views idiom and the empirical $N^{\ast}\!=\!1$ on medium/large databases shows the idiom translates almost verbatim. Whether \Sgcf{} extends to other database-side context layers (constraints, lineage, access policies, query logs) is left as explicit future work rather than claimed here. Together, these takeaways argue that database-side context engineering deserves first-class status alongside prompt design and linker architecture, and the offline/online split makes the resulting compressors immediately deployable atop existing Text-to-SQL stacks without retraining.

\textbf{Conclusion.} Database context compression, grounded in \Sgcf{} and realised in \DbCC{}, raises strict recall on enterprise-scale databases from zero to the mid-fifties / low-sixties (LLM-dependent) while cutting input tokens by roughly two orders of magnitude, with benefits that compose additively over stronger LLMs and existing schema-linking pipelines.

\bibliographystyle{IEEEtran}
\bibliography{references}

\end{document}